\documentclass[singleside]{article}
\usepackage{fleqn,espcrc2,times}
\usepackage{graphicx}
\usepackage{latexsym}
\usepackage{amssymb}
\DeclareGraphicsExtensions{.eps,.ps}
\title{Nodal Quasiparticles versus Phase Fluctuations in High
 \textit{T}$_{\bf c}$ Superconductors: An Intermediate Scenario}
  \author{Qijin Chen, Ioan Kosztin, and K. Levin%
  \address{The James Franck Institute, The University of Chicago, 5640
    South Ellis Avenue, Chicago, Illinois 60637}}
\begin{document}
%----------------------------------------------------------------

\begin{abstract}
  We explore a BCS Bose Einstein crossover scenario for $0 \leq T \leq T_c$
  and its implications for the superfluid density and specific heat. The low
  lying excitations consist of nodal (fermionic) quasi-particles as well as
  excited (bosonic) pair states.  Semi-quantitative comparison with cuprate
  data is reasonable, with no compelling indications for Landau parameter
  effects.
\vskip -0.5mm
\end{abstract} 
%}
\maketitle

\section{Introduction}

Within the pseudogap regime of the cuprates, it has been widely argued that
the excitations of the superconducting state are predominantly
fermionic\cite{Lee} or predominantly bosonic\cite{Emery} in character.  We
have pointed out\cite{chen} that there is a third scenario which is no less
likely and which, at the least, needs to be considered on an equal footing.
This third scenario emerges when one studies the BCS Bose-Einstein
condensation (BEC) crossover picture for $ 0 \leq T \leq T_c$.  At weak
coupling (when the coherence length $\xi$ is large) the excitations are
dominantly fermionic, and at very strong coupling (small $\xi$) they are
dominantly bosonic.  At intermediate coupling, which is likely to be
appropriate to the cuprates, the excitations are of mixed character. In this
paper we discuss the implications of this third scenario, in the context of
semi-quantitative comparisons with experiment, for the superfluid density
and specific heat.

There have been a number of recent papers which have presented similar
comparisons within the context of the `` fermionic" or nodal quasi-particle
picture, extended, however, to a Fermi liquid based
interpretation\cite{Millis,Norman,Taillefer,Cambridge}.  This differs
conceptually from the original formulation of Lee and Wen\cite{Lee} because
the important Mott insulator constraint ($\omega_p^2 \approx x $) is
enforced via a hole concentration ($x$) dependent Landau parameter $F_1 ^s$
which then constrains the penetration depth $\lambda (x)$ via $ d \lambda
^{-2} / dT \approx x^2$.  These $x$ dependences are considerably different
from those of the spin-charge separation approach\cite{Lee} which is based
on the presumption that $ d \lambda ^{-2} / dT $ is $x$ independent.

An underlying philosophy of the present paper is that fundamental features
of the pseudogap phase should be accommodated at the outset, before
including Landau parameter effects.  Our viewpoint is different from the
phenomenology of Lee and Wen\cite{Lee} because here $T_c$ is associated with
\textit{both} nodal quasi-particles and bosonic pair excitations. Stated
alternatively, in the BCS-BEC crossover picture there is an important
distinction between the order parameter $\Delta_{sc}$ and excitation gap
$\Delta$ at \textit{all} temperatures.

%\suppressfloats
\setlength{\parskip}{0in}

This approach leads to a new mean field-like 
theory\cite{chen} which incorporates
(i) the usual BCS equation for $\Delta(T)$ and for (ii) the chemical
potential $\mu (T)$, along with a third new equation for 
\linebreak

\begin{figure}[hbt]
%\vskip -8mm
\vspace*{1.5ex}
\centerline{\includegraphics[width=2.72in, clip]{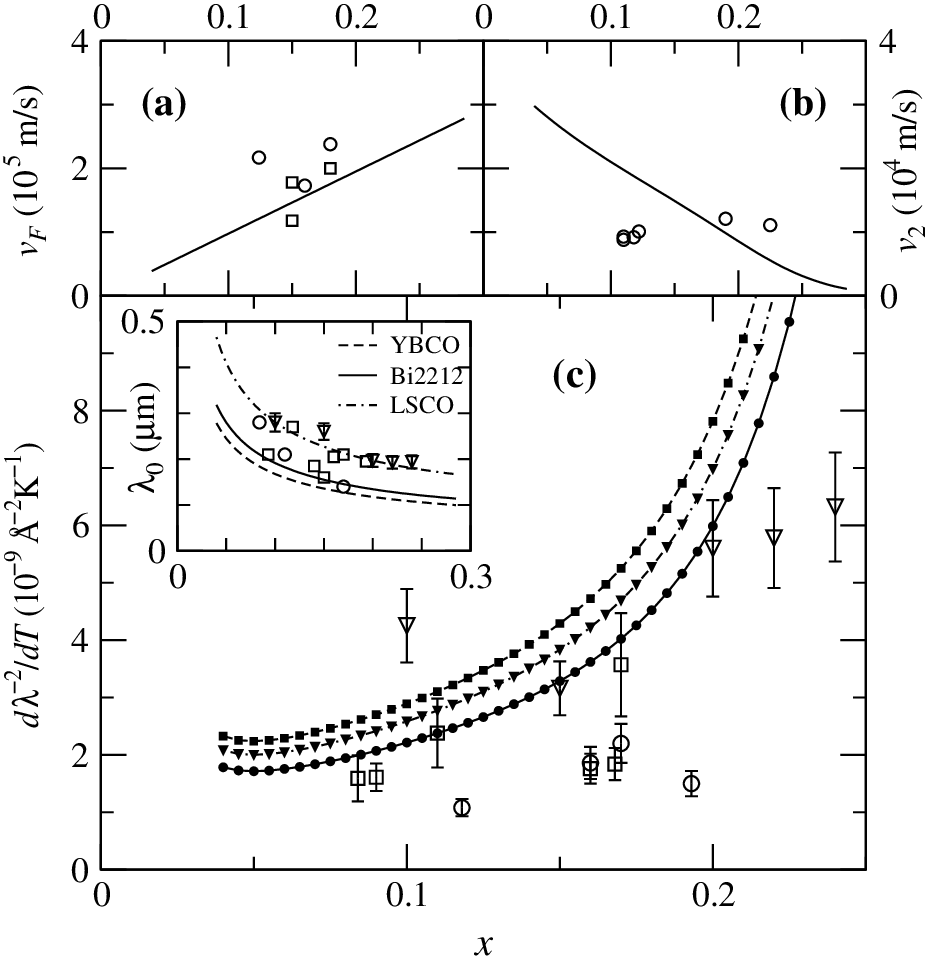}}
\vskip -1cm
\caption{\small Input parameters (1a,1b) to  
  $d\lambda^{-2}/dT$ plotted in 1c, for various cuprates indicated by the
  legend in Fig. 2a.  Inset in 1c shows $\lambda_0$ vs doping $x$. Data on
  $v_2$ from Ref.~\protect\cite{Norman}, $v_F$ from ({\large $\circ$})
  Refs.~\protect\cite{Valla,Zhou,Waldmann,LeeSF} ({\footnotesize
    $\Box$})\protect\cite{Lee,Krishana}, others ({\large $\circ$}, Bi2212)
  Refs.~\protect\cite{Waldmann,LeeSF}; ({\footnotesize $\Box$}, YBCO)
  \protect\cite{Hardy,Panagopoulos98}; ({\footnotesize $\triangledown$}, LSCO)
  \protect\cite{Panagopoulos99}.} \vskip -7mm
\end{figure}

\vskip -2ex \noindent 
(iii) $\Delta^2 - \Delta_{sc}^2$ which is related to
the number of thermally excited pair (bosonic) excitations.  The first two
equations enforce an underlying fermionic constraint so that the bosons of
the strong coupling limit \textit{are different} from those of a true boson
system, such as He$^4$.  To include the Mott insulator constraint the Fermi
velocity $v_F$ must then be $x$-dependent, as shown in Figure 1a (along with
experimental data), in a way which \textit{directly} reflects the
$x$-dependence of $\lambda (T=0) = \lambda_o$, shown in the inset to
Figure 1c. Here the parameters were chosen to give a reasonable fit to the
measured phase diagram for the YBaCuO system\cite{chen}.

\begin{figure}
\vspace*{2ex}
\centerline{\includegraphics[width=2.75in, clip]{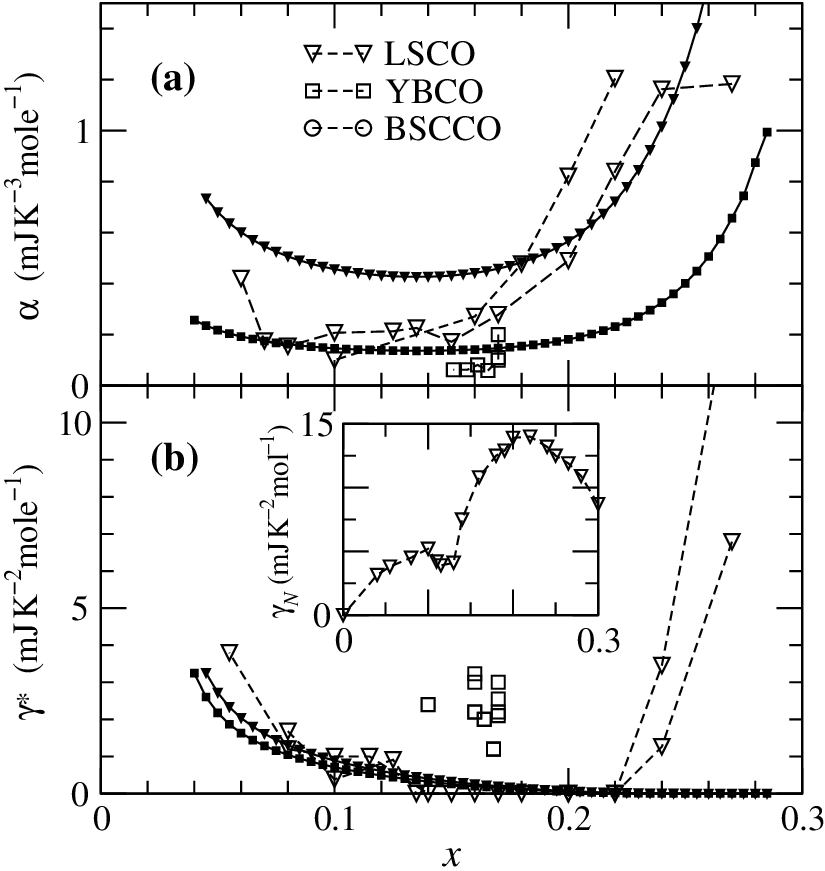}}
\vskip -9mm
\caption{\small Quadratic (2a) and linear (2b) $T$ contributions to $C_v$
  in various cuprates. Data are from ({\footnotesize $\triangledown$})
  Refs.~\protect\cite{Loram,Momono} and ({\footnotesize $\Box$})
  Refs.~\protect\cite{Moler,Junod,Wright,Fisher}. Inset in 2b shows normal
  state result from Ref.~\protect\cite{Momono}. Compare open (experiment)
  with corresponding filled (theory) symbols. (per mol of formula units)} \vskip -5mm
\end{figure}
 
\vspace*{-2mm}
\section{Penetration Depth and Specific Heat}
\vskip -1mm

In order to calculate the penetration depth and specific heat within the
present approach, the fermionic contributions must be quantified, just as in
the nodal quasi-particle picture. This contribution is accompanied by an
additive bosonic component\cite{chen,chen2}.  So as not to complicate the
logic, we compute the former, by taking the second velocity
contribution\cite{Lee} $v_2$ as given by a perfect $d$-wave model; thus the
$x$ dependence of $v_2$ entirely reflects that of $\Delta(x)$, as shown in
Fig. 1b, which is in slight disagreement with the data indicated in
1b\cite{Norman}. The resulting values for the inverse squared penetration
depth are plotted in Fig.~1c, along with a collection of experimental
data.  The predicted increase at large $x$ is a reflection of the behavior
of $\Delta (x)$. Given the spread in the data for all quantities indicated
in Fig.~1, it would appear that there are no obvious inconsistencies.

In Fig.~2 we show the $x$ dependent coefficient of the quadratic term (2a)
in the specific heat $C_v = \gamma^* T + \alpha T^2$, along with the linear
term (2b).  The first of these reflects the fermionic quasi-particle
contribution (which depends on $v_2$ and $v_F$) and the second derives
purely from the bosonic contribution.  Also indicated are a collection of
experimental data on three different cuprates. We know of no other intrinsic
origin for this $\gamma^*$ term, which despite its widespread presence is
usually attributed to extrinsic effects.

The upturn at large $x$ in the $\gamma^*$ data, is of no concern, since it
is a reflection of the normal state behavior (shown more completely in the
inset). For overdoped samples, at the lowest $ T_c \approx 0$,
extrinsic,e.g., paramagnetic impurity, effects make it difficult to observe
$\alpha$ and $\gamma^*$ in the intrinsic superconducting state.

%\bibliographystyle{prsty} 
%\bibliography{/home/qchen/bibinputs/M2S}

\end{document}